\documentstyle[12pt]{article}

\begin{document}
\voffset 1cm
\renewcommand{\thesection}{\arabic{section}.}
\newcommand{\tq}{top quark }
\newcommand{\nn}{\nonumber}
\newcommand{\nl}{\newline}
\newcommand{\gp}{g^{\prime} }
\newcommand{\gpp}{g^{\prime\prime} }
\newcommand{\be}{\begin{equation}}
\newcommand{\ee}{\end{equation}}
\newcommand{\bea}{\begin{eqnarray}}
\newcommand{\eea}{\end{eqnarray}}
\newcommand{\lb}{\label}
\newcommand{\pr}{\prime}
\newcommand{\al}{\alpha}
\newcommand{\ga}{\gamma}
\newcommand{\de}{\delta}
\newcommand{\De}{\Delta}
\newcommand{\eps}{\epsilon}
\newcommand{\ti}{\times}
\newcommand{\half}{\frac{1}{2}}
\newcommand{\ra}{\rightarrow}
\newcommand{\lra}{\longrightarrow}
\newcommand{\mi}{\mbox{i}}
\newcommand{\mtr}{\mbox{Tr}\,}
\newcommand{\ttttil}{\tilde{T}}
\newcommand{\ttpmtil}{\tilde{T}_{\pm}}
\newcommand{\ttztil}{\tilde{T}_{0}}
\newcommand{\ttptil}{\tilde{T}_{+}}
\newcommand{\ttmtil}{\tilde{T}_{-}}
\newcommand{\tytil}{\tilde{Y}}
\newcommand{\ttt}{\bar{T}}
\newcommand{\ttpm}{\bar{T}_{\pm}}
\newcommand{\ttz}{\bar{T}_{0}}
\newcommand{\tti}{\bar{T}_{i}}
\newcommand{\ttj}{\bar{T}_{j}}
\newcommand{\ttk}{\bar{T}_{k}}
\newcommand{\ttdrei}{\bar{T}_{3}}
\newcommand{\ttzwei}{\bar{T}_{2}}
\newcommand{\tteins}{\bar{T}_{1}}
\newcommand{\ttdreitil}{\tilde{T}_{3}}
\newcommand{\ttzweitil}{\tilde{T}_{2}}
\newcommand{\tteinstil}{\tilde{T}_{1}}
\newcommand{\ttp}{\bar{T}_{+}}
\newcommand{\ttm}{\bar{T}_{-}}
\newcommand{\ty}{\bar{Y}}
\newcommand{\tgr}{\bar{g}_R}
\newcommand{\dd}{1-2\De}
\newcommand{\lr}{SU(2)_L \ti SU(2)_R \ti U(1)_{B-L}}
\newcommand{\sm}{SU(2)_L \ti U(1)_Y}
\renewcommand{\thefootnote}{\fnsymbol{footnote}}
\begin{titlepage}
\begin{quote} \raggedleft  {\bf DESY 94-129} \end{quote}
\begin{center}
{\bf

Transition from
SU(2)$_{\mbox{\small L}} \ti  $
SU(2)$_{\mbox{\small R}} \ti$ U(1)$_{\mbox{\small B-L}}$
Representation
to SU(2)$_{\mbox{\small L}} \ti$ U(1)$_{\mbox{\small Y}}$
by q-Deformation
and the \\ Corresponding Classical
Breaking Term of Chiral U(2)}
\end{center}
\vspace{1cm}
\begin{center} {R. B\"onisch}
\\
Brookhaven National Laboratory, Upton, New York 11973;
\\ DESY-IfH, Platanenallee 6, D-15738 Zeuthen (since April 94)
\end{center}
\begin{abstract}
The minimal Standard Model exhibits a nontrivial
chiral $U(2)$ symmetry
if the {\sc vev} and the hypercharge
splitting $\De = (y_R^u-y_R^d)/2$ of right-handed
leptons (quarks) in a family vanish and
$Q=T_0 + Y$ independently in each helicity sector.
As a generalization,
we start with $SU(2)_L \ti SU(2)_R \ti U(1)_{(B-L)}$
and introduce $\De$ as a continuous parameter which is a measure of
explicit symmetry breakdown.
Values $0 \leq \De \leq 1/2 $ take
the neutral generator of the
isospin-$\frac{1}{2}$ representation to the singlet representation,
i.e. `deformes' the LR representation into the minimal Standard one.
The corresponding classical $O(3)$-breaking term is a magnetic field
perpendicular to the $x_3$-axis.
A simple mapping on the fundamental
Drinfeld-Jimbo
$q$-deformed $SU(2)$ representation is given.
\end {abstract}
\end{titlepage}
\renewcommand{\thefootnote}{\fnsymbol{footnote}}
\section{Introduction}
The very underlying symmetry of a system is the largest nontrivial
symmetry that can be obtained in any limiting set of values of the
parameters of the system. This symmetry refers to the minimal
structure that occurs in the symmetric phase and
may be broken at an arbitrary set of values
of its parameters.
The electroweak Standard Model (SM) of one fermion family exhibits a
chiral $U(2)$,
\be
U(2)_L \times U(2)_R = SU(2)_L \times U(1)_{Y_L}
\times SU(2)_R \times U(1)_{Y_R}.
\lb{chiralu2}\ee
The $U(1)$ charges are arranged
such that the electric charge is given as a combination
of the generators in the Cartan subalgebra
of each chirality sector:\footnote{
Projectors $\frac{1}{2}(1 \pm \ga_5)$ fix normalization
of $Q$ in $J_{em}=\bar{\Psi} \ga^\mu Q \Psi$ such that $Q_L =Q_R=Q
=$diag$(0,-1)$ [diag$ (2/3, -1/3)]$ for leptons [quarks] and
we use $T_0=\frac{1}{2} \tau_3$,
$T_\pm = \frac{1}{\sqrt{2}} (\tau_1 \pm \mi \tau_2)$,
where $\tau_i$ are Pauli
matrices normalized to $\tau^2 =1$.
These choices avoid further factors 1/2 in quantum numbers.}
\be Q_x = T_{0x} + Y_x; \hspace{2cm} x=L,R \lb{qlr}.\ee
The $SU(2)_R$ factor
is hidden in the fermion representation of the minimal SM:
$T_{R} \equiv 0$.
Therefore, $T_{0R} \equiv 0$ and
we have a splitting
\be \Delta \equiv \frac{y^u_R - y^d_R}{2} \lb{delta}\ee
of hypercharges $y^{u,d}$
in the right-handed sector of potential isospin-$\frac{1}{2}$ components
with a charge matrix
\be Y_R = Q= \left( \begin{array} {cc} y^u_R & 0 \\ 0 & y^d_R \end{array}
\right). \lb{yrq}\ee

Clearly, eqn.
 (\ref{chiralu2}) and  (\ref{qlr}) are not unique
as any hidden (trivial) factor can in principle be added to the standard
$SU(2) \ti U(1)$
symmetry and to $Q$ and only
the requirement of eq.  (\ref{qlr}) makes the
$SU(2)_R$ definitely appear in eq.  (\ref{chiralu2}).
The search for the origin of electroweak symmetry breakdown
is the search for the underlying symmetry and for the parameters
which are nonzero in the broken phase.

One crucial quantity in the symmetry breakdown is the splitting
\linebreak[4]
$\De_g = g_u-g_d$
of Yukawa-couplings $g_{u,d}$ within an isospin doublet.
Finite $\De_g$ breaks
$SU(2)$ down to $U(1)$, but $\De_g$
can hardly be regarded independent from either the {\sc vev}
$v$ or $\De$:
the renormalization procedure yields Yukawa-couplings
with $U(1)$ corrections such that even
possibly degenerated tree-level couplings are
always infected with
$SU(2)$-violating self-energies. Therefore $\De_g$ will not vanish
in the physical lagrangian at an arbitrary scale.
Masses $m$ indeed show a numerical connection
to charge, which
has recently been put into an empirical formula for
$m_i$ across the families $i=1,2,3$
\cite{Sirlin}: \be \frac{m_2}{m_1}=3 \left( \frac{m_3}{m_2}
\right)^{\frac{3}{2}|Q|}. \lb{mi}\ee
The reason for this behaviour with $Q$ is unknown
and it can actually
also be hypercharge that is involved, eq.  (\ref{yrq}).
The factor 3/2
may however point to some deviation from $Q$ in the origin of
$SU(2)$ violations.
Various ways to connect $\De$ and $\De_g$
are possible:
pure radiative effects \cite{Fritzsch,u1mass} or
direct strong fermionic coupling to hypercharge currents \cite{Bonisch}
can be used.
The existence of fermion mixing however favors gauge
eigenstates, not mass eigenstates to be important in the
true mechanism.
Eq.  (\ref{mi}) states the importance of the $U(1)$ factor for the
mass generation mechanism: it determines horizontal as well as
vertical mass gaps in the standard spectrum. It also tells us that
the relevant quantity for isospin violation is already known and does
not have to come from beyond the SM.
At the same time, as Yukawa-couplings are
only meaningful once chirality is broken,
the responsible interaction may not show maximal parity violation
\cite{Fritzsch2}, but must possess a (nontrivial) diagonal subgroup.

The largest nontrivial chiral symmetry we get
as a limit of the standard $SU(2)_L \ti U(1)_Y$ is  (\ref{chiralu2}):
We require eq.  (\ref{qlr})
and take
\be v \ra 0, \hspace{2cm} \De \ra 0. \lb{limit}\ee
This yields  (\ref{chiralu2}) just in the representation of the
much explored left-right symmetric model (LR);
$SU(2)_R$ is not hidden and $ Y_L =Y_R = \frac{1}{2} (B-L)$
\cite{Senjanovic+Mohapatra}.
Parity will be broken
spontaneously by $SU(2)$ couplings $g_L \neq g_R$
or {\sc vev}'s $v_L \neq v_R$ and should be restored in the symmetric
phase at higher energies.
At $\De =0$ however, $Y$ cannot introduce a
vertical mass gap in eq.  (\ref{mi}) and
chirality breaking is also suppressed
if one chirality sector is hidden.
{}From this point of view, more general values of $\De$ are interesting.

If the underlying symmetry of the SM is broken explicitly,
it is interesting to ask for a physical term for this breaking
in the lagrangian also irrespective of any details of the mass spectrum.
If e.g. the SM has a nontrivial
$U(2)_L \ti U(2)_R$ symmetric limit, but parity
is broken not only in low energy states,
a breaking term should be the corresponding physical sector that stands for
initial symmetry breakdown.

The purpose of this paper is to construct
a class of algebras that includes the LR representation as
one limit and the minimal SM representation
as the other and to calculate the classical symmetry breaking term.
In section 2 we generalize the $SU(2)$ algebra by introducing
$\De$ as a continuous parameter and show that the
corresponding $O(3)$-breaking term is a magnetic field perpendicular
to the $x_3$-axis direction and proportional to
$\De$. In section 3 we show the connection
to the well-known  $SU(2)_q$ by Drinfeld and Jimbo
in a simple mapping between $\De$- and $q$-representations.
Section 4 consists of a remark on the freedom of introducing
a further parameter in the solution of the classical system and
section 5 is the summary.
\section{General $\De$ and the Classical $SU(2)_R$ Breaking Term}
We consider the well-known Left-Right Model \cite{Senjanovic+Mohapatra}
and
the Standard Model with no Higgs sector coupled to the fermion-boson content.
The parameter that connects LR and SM representations
$\lr$ and $\sm$ is the
hypercharge-splitting eq. (\ref{delta}) taking continuous values
$$0 \leq \De \leq \frac{1}{2}.$$

To break the symmetry, we perform a
non-orthogonal $GL(2)$ transformation in the right-handed
Cartan subalgebra of the LR model to obtain
a generalized basis
$(\ty, \ttz)$, which only satisfies eq.  (\ref{qlr}):
\bea \left( \begin{array}{c} \ty \\ \ttz \end{array} \right)
= \left( \begin{array}{cl} 1 & 2\De \\ 0 & 1-2\De \end{array} \right)
\left( \begin{array}{c} \frac{1}{2} (B-L) \\ T_0 \end{array} \right).
\lb{bmix}\eea
The LR model chooses $\De = 0$ while in the SM $\De =1/2$.
$(B-L)$ is
anomaly free and so is $\ty$: the relevant sums of triangle diagrams
involving 1 to 3 external $U(1)$ gauge fields are
proportional to
$ \mbox{Tr} (\{ \ttt_i, \ttt_j\} \ty)$, $\mbox{Tr}(\ttt_i \ty^2)$ and
$\mtr({\ty^3})$. Using eq.  (\ref{qlr}) isolates pure vectors, which
have powers only of $Q$, but by eq. (\ref{bmix})
the cancelation is seen to work within each family:
\bea & &\mbox{Tr}\, (T_0^3)=0; \nonumber \\
& &\mtr [T_0^2(B-L)] = \mtr (B-L)
=\frac{N_q \cdot N_c}{3} - N_l; \nonumber \\
& &\mtr [T_0 (B-L)^2 ] \sim \mtr (T_0) =0; \nonumber \\
& &\mtr
[(B-L)^3] = N_q N_c \left (\frac{1}{3} \right)^3 - N_l, \lb{anomal}\eea
where $N_q$ $(N_l)$ is the number of quarks (leptons).
The left-handed sector is identical to the SM one and, like in the
SM,
the abelian factor is gauged only in the vector subgroup,
so that the last of eqn.  (\ref{anomal}) does not contribute.

The symmetry under consideration is now
\be SU(2)_L \ti U(1)_{(B-L)_L} \ti \left[ SU(2)_R \ti U(1)_{(B-L)_R}
\right]_\De. \ee
$\ttz$ is fixed and the question is how to close
the $[SU(2)_R]_\De$ algebra.
$[SU(2)_R]_\De$ should contain the charged sector of the LR model as $\De \ra
0$, but is otherwise unspecified.

In a simple isotropic renormalization of $SU(2)$ generators
\be \ttz = (1-2\De) T_0, \hspace{2cm} \ttpm = \sqrt{1 - 2 \De} T_\pm, \ee
all $\De$-dependence could be absorbed into the (primordial)
gauge coupling,
\be g_R \ra \bar{g}_R(\De) \equiv (1-2\De)g_R. \ee
$SU(2)_R$ gauge transformations would then read
\be D_\mu \Psi_R \ra
\mbox{exp}[\mi \tgr T_j \omega(x)^j_R] D_\mu \Psi_R, \ee
\be \bar{\Psi}_R \ra \bar{\Psi}_R \,\,
\mbox{exp}[-\mi \tgr T_j \omega(x)^j_R],
\ee
where $D_\mu=\partial_\mu + \mi \tgr T_j W^j_\mu$ and we have
\be [\ty, T_0]=0, \hspace{2cm} [\ty, T_\pm]=\pm 2\De T_\pm. \lb{tybar}\ee
As long as the $B_\mu$ gauge field behaves trivially
under $SU(2)_R$,
the hypercharge interaction term
$  J^\mu_{\ty_R}B_\mu =  \bar{\Psi}_R
[\frac{1}{2}(B-L)+2\De T_0]\ga^\mu \Psi_RB_\mu $
is not invariant:
\be \mi \bar{\Psi}_R \ga^\mu \ty \Psi_R B_\mu \ra
\mi \bar{\Psi}_R \ga^\mu \ty \Psi_R B_\mu
- \tgr \bar{\Psi}_R \ga^\mu [\ty, T_j\omega^j] \Psi_R B_\mu, \ee
because of the $T_0$ term in $\ty$. It is proportional to
$\tgr \cdot \De$, which vanishes at both
$\De =0$ and $\De = 1/2$, i.e. in LR and SM representations, and
is maximal at $\De = 1/4$.
The first of eqn. (\ref{tybar}) ensures the $U(1)$ to survive in
the breakdown
\be U(2)_R = SU(2)_R \ti U(1)_{Y_R} \stackrel{\De \neq 0}{\lra} U(1)_{Y_R} \ti
U(1)_{T_0}.
\lb{rbreak}\ee

We are interested in explicit breaking of chiral $U(2)$ to
the SM $\sm$
and a corresponding physical term in the Lagrangian.
Therefore let us now consider a classical system and
generalize
the ordinary $SU(2)$-algebra with commutation relations
\be [T_0,T_\pm]=\pm T_\pm, \hspace{2cm}
[ T_+, T_-]=T_0 \lb{su2}\ee
and Casimir
\be C = 2 \,T_+\, T_- + T_0(T_0-1)\,= 2\,T_-\, T_+ +T_0(T_0+1) \equiv
T\,(T+1). \lb{casi}\ee
Eq. (\ref{bmix}) together with leaving $T_\pm$ unchanged
can be taken as a deforming map on the fundamental
$SU(2)$ representation \cite{Curtright+Zachos} and
closing $\ttz$ with $T_\pm$ gives
\be  [\ttz, T_\pm] =\pm  (1-2\De) \,\, T_\pm,
\hspace{1.2cm} [T_+,\, T_-]= (1-2\De)^{-1} \ttz.
\lb{quom}\ee
This corresponds to a map
\be T_3 \ra
\ttdrei = (1-2\De) T_3,\hspace{1.2cm} T_{1,2}\ra \bar{T}_{1,2} = T_{1,2} \ee
into the Cartesian basis of $O(3)_\De$ and
\be [ \tti, \, \ttj] = \bar{\eps}_{ijk} \ttk, \ee
\be \bar{\eps}_{ijk} = \left\{ \begin{array}{ll}
\pm (1-2\De) ^{-1}  & \mbox{for $i\,(j)=1\,(2),$  $k=3$} \\
\pm (1-2\De) &  \mbox{for $i$ or $j=3,$  $k=1$ or $2$} \\
0 & \mbox{else} \end{array} \right. \lb{bareps}\ee
and $\bar{\eps}_{ijk} \ra \eps_{ijk}$ with $\De \ra 0$.
Using $\bar{\eps}_{ijk}$ from eq. (\ref{bareps}) to write down
$\ttt $ in $O(3)_\De$ is equivalent to
rescale $O(3)$ vectors
\be {\bf x} \ra {\bf x}^\pr
=(x_1,\,x_2,\,x_3^\pr=\{1-2\De\}x_3) \lb{rescale}\ee
with the result
\bea
T_1^\pr&=&\, {\bf e}_1^\pr  (x_2\,p_3-\,x_3\,p_2)= \,T_1 \nn \\
T_2^\pr&=&\, {\bf e}_2^\pr  (x_3\,p_1-\,x_1\,p_3)= \,T_2 \nn \\
T_3^\pr&=&\, {\bf e}_3^\pr\,(x_1\,p_2-\,x_2\,p_1) = (1-2\De)\,T_3.
\eea
For a classical free particle, eq.  (\ref{rescale}) produces a scalar
potential $U$,
\bea L &\ra& L^\pr = \frac{1}{2m}p^{\pr2} = L - U \lb{lu}\\
U&=&-\frac{2\De}{m}p_3^2 +..., \eea
which contains a vector potential $\bf A$. In the present case we have
\be U= - {\bf A} \cdot{\bf v}\ee
and ${\bf A} \equiv (0,\,0,\,A_3)$ corresponds  to a magnetic field
\be {\bf B} = {\bf e}_1 \partial_2 \, A_3-{\bf e}_2 \partial_1 \, A_3.\ee
Note that for a constant $\bf A$ the deformation parameter becomes
momentum dependent,
\be  A_3 = 2\,{ \De}_c =2 \De p_3,\lb{Dep}\ee
where ${ \De}_c = \De p$ enters eq.
 (\ref{quom}) and subsequent definitions of $O(3)_{\De_c}$ quantities.
The classical free $\De$-particle is thus a particle in a magnetic field
$\bf B$ which is in the $(x_1,x_2)$-plane, while
$\bf A$
breaks $O(3)$ explicitly and diverges with $\De \ra \half$.
The limits are now:
\be \lr \hspace{1cm} \mbox{at} \hspace{1cm} \De = 0, \ee
\be \sm \hspace{.3cm} \mbox{plus potential } U
\hspace{.5cm} \mbox{at} \hspace{.5cm}\De = \half,
\lb{smlimit}\ee where $U$ is the desired augmentation of $L$ in eq.
(\ref{lu}).

A general vector potential with components $A_i=2\De_i p_i=2\De_{i}$
yields
\be L^\pr=\frac{1}{2m} (p^2+4{\bf \De}{\bf p}) \lb{gen1}\ee
and corresponds to a rescaling
\be  x_i \ra  x^\pr_i=(1-2\De_{i})x_i, \lb{gen2}\ee
where we dropped the index $c$.
\section{Connection with q-Deformation}
The most studied generalization of $SU(2)$ is the Drinfeld-Jimbo $SU(2)_q$
algebra, given by
\be [\ttz,\,\ttpm]=\pm \ttpm, \hspace{2cm}
[ \ttp, \ttm]= [\ttz]_{q^2}, \lb{su2q}\ee
where $[x]_{q^2} \equiv (q^{2x}-q^{-2x}/(q^2-q^{-2})$ and
$q \ra 1$ yields $SU(2)$. The deformation  (\ref{su2q})
succesfully describes e.g. the anisotropic Heisenberg model.
Comultiplication rules and many details of its representation theory
are known \cite{Ruiz-Altaba}.

Following Curtright and Zachos \cite{Curtright+Zachos},
we can find expressions $\ttpm$ as functions of $SU(2)$ generators
with the Ansatz
\bea
\ttp \, \ttm &\equiv&  \frac{1}{q+1/q} \cdot
\left\{ [T]_q[T +1]_q\,-\,[\ttz]_q[\ttz -1]_q \right\}
\nn \\
\ttm \, \ttp &\equiv&  \frac{1}{q+1/q} \cdot
\left\{ [T]_q[T +1]_q\,-\,[\ttz]_q[\ttz +1]_q \right\}. \lb{ttq}\eea
The Casimir is
\be C_q = 2 \,T_+\, T_- + [\ttz]_q[\ttz-1]_q\,
= 2\,T_-\, T_+ + [\ttz]_q[\ttz+1]_q\,=[T]_q[T+1]_q. \lb{casiq}\ee
{}From eq.  (\ref{casi}) we have $2 \, T_+\, T_- =  (T+T_0)\,(1+T-T_0)$ and
$2 \, T_-\, T_+ =  (T-T_0)\,(1+T+T_0)$ and with the $q$-analogues
and eq. (\ref{ttq}) we arrive at
\bea
\ttp &=&  \sqrt{ \frac{2}{q+1/q}\cdot
\frac{[T+T_0]_q\,[1+T-T_0]_q}{(T+T_0)\,(1+T-T_0)}} \, T_+, \nn\\
\ttm &=&  \sqrt{ \frac{2}{q+1/q}\cdot
\frac{[T-T_0]_q\,[1+T+T_0]_q}{(T-T_0)\,(1+T+T_0)}} \, T_-, \lb{ttpm}\eea
and, from eq. (\ref{su2q}), $\ttz =T_0$.
Inserting the $T=\half$ representation into eq. (\ref{ttpm}) gives
\be \ttpm=\al T_\pm,  \hspace{1cm}
\al =\sqrt{2/(q+1/q)}. \lb{alpha}\ee
In the cartesian basis we therefore have
\be \ttdrei = T_3,\hspace{1cm} \bar{T}_{1,2} = \al T_{1,2} \ee
and the commutators
\be [\tteins,\,\ttzwei]=\mi\al^2 T_3,
\hspace{.5cm} [T_3,\,\tteins]=\mi \ttzwei,
\hspace{.5cm} [\ttzwei,\,T_3]=\mi \tteins, \ee
correspond to a rescaling
\be {\bf x} \ra {\bf x}^\pr =(\al\,x_1,\al\,x_2,\,x_3).\lb{qdef}\ee

We identify
\be 1-2\De=\sqrt{\frac{2}{q+1/q}}  \lb{conn} \ee
and eqn.  (\ref{gen1}) and  (\ref{gen2}) yield a potential ${\bf A}=(A_1,\,A_2,
\,0)$, which corresponds to a field $\bf B$ in $x_3$-direction.
This agrees with the interpretation in \cite{Martin-Delgado}. Note that
the classical limit $q \ra 1$ corresponds to $\De \ra 0$ in eq.  (\ref{conn}).
The second limit eq. (\ref{smlimit}) $\De \ra \half$ corresponds to $q \ra
\infty$. Note the symmetry $q \ra q^{-1}$.

We rotate vectors $\bf x$ to ${\bf \hat{x}}=R \,{\bf x}$, where
\be R= \left( \begin{array}{ccc} r^2 & -r^2 & -r \\ -r^2 & r^2 & -r \\
r & r& 0 \end{array} \right) \lb{r}\ee
and $r= \sin (\pi/4)$. After applying
the map eq. (\ref{qdef}) they take the form
\bea
\hat{x}_1^\pr &=& \al r^2 x_1 - \al r^2 x_2 - r x_3 \nn \\
\hat{x}_2^\pr &=& -\al r^2 x_1 + \al r^2 x_2 - r x_3 \nn \\
\hat{x}_3^\pr &=& \al r (x_1 + x_2) = \al (\hat{x}_1 + \hat{x}_2) \eea
and with eq.
(\ref{conn}) the third component satisfies eq. (\ref{bmix}) again.

We note that only $\ttz$ is fixed in eq. (\ref{bmix}) while $\ttpm$
are only required to close the algebra and yield the classical limit.
There is also some freedom in normalizing the structure constants
$\bar{\eps}$ or commutators, namely $[\ttp, \ttm]=\half [2 \ttz]_q$ does
not exhibit the deformation in the generators of the $T=\half$ representation
\cite{Curtright+Zachos}.
\section{Change of Topology}
We want to remark that there is actually
more freedom in the dynamics of the classical particle
than considered in sections 1 and 2.
Eq. (\ref{bmix}) fixes the generators of the Cartan subalgebra
while `charged operators' $\ttpm$ are only required to close $SU(2)_\De$ and
yield ordinary $SU(2)$ when $\De \ra 0$.
The effect of the symmetry breaking parameter being nonzero can
in principle be more drastic than
$GL(2)$ transformations on the ordinary representations, namely the topology
might be altered in the following sense.

Eq. (\ref{alpha}) directly shows that $\ttpm$ create the vacuum out of
lowest and heighest weight states: the ordinary normalization of states
$\ttpm |t,t_0 \rangle =$
\linebreak[4] $\sqrt{(t\mp t_o)(t \pm t_o +1)}|t,t_0 \pm 1
\rangle$ has become
\be\ttpm |t,t_0 \rangle = \al\sqrt{[t\mp t_o]_q[t \pm t_o +1]_q}|t,t_0 \pm 1
\rangle, \ee
\cite{finrep}, and $\ttpm |t,t_0 \pm 1 \rangle =0$.

The Ansatz eq. (\ref{ttpm}) can however be relaxed by adding
to $\ttp \ttm$ and $\ttm \ttp$ a term
$f(q)=a(q)[\ttp + \ttm + a(q)]$, i.e. adding $a(q)$ to $\ttp$ and $\ttm$.
In the difference eq. (\ref{su2q}), $f(q)$ vanishes.
For $a(q)$ a number, the new representation would for example read
\bea
\ttptil &=& \ttp + a(q) = \left( \begin{array}{cc} a & \al \\ 0 & a
\end{array} \right) \nn\\
\ttmtil &=& \ttm + a(q) = \left( \begin{array}{cc} a & 0 \\ \al & a
\end{array} \right). \lb{anom} \eea
If $\ttpm + a = \tteinstil \pm \mi \ttzweitil$ one sees that
this would correspond to
\be {\bf x} \ra {\bf x}^\pr = {\bf x}+ a \ee
and $a$ is not a parameter of $O(3)$-geometry, but gives a coordinate
independent term (constraint) in the equations of motion.

In a gauge theory, one would expect anomalies to appear in this case and
if we naively enter the sum of triangle diagrams with the `anomalous
couplings'
eq. (\ref{anom}), we get a contribution proportional to $\mbox{Tr}
\tilde{T}_i=2a$.
\section{Summary}
On the basis of chiral $U(2)$,
we defined a continuous transition from the $\lr$ representation
of the Left-Right model to the $\sm$ representation of the minimal Standard
Model. The transition parameter $\De$ is the splitting of hypercharges of
right-handed up- and down-type fermions
in the $T=\half$ representation. Classically, $\De$ is proportional to
a (iso)magnetic
field which vanishes in the LR and diverges in the limit of the
SM.
$\De$ is therefore a measure of explicit breaking of chiral $U(2)$.
The same happens in $q$-deformation such that
we get \linebreak[4]
$\De = \half \left[ 1- \sqrt{2/(q+q^{-1})} \right]$.
\vspace{.5cm}\nl
{\bf Acknowledgement}\nl
The author wishes to thank
the Humboldt-Universit\"at Berlin and the
CERN Theory Divison very much for invitations and hospitality and
J. Erler and \linebreak[4]
S. Gavin for comments during the work,
and C. Aneziris, D. L\"ust and \linebreak[4] G. Weigt for
discussing the manuscript.
\pagebreak[4]

\end{document}